\documentclass[st,twocolumn]{jpsj3}
\usepackage{txfonts}
\usepackage{amsmath,braket}
\usepackage{graphicx,color}

\title{Geometric Effects on Tunneling in Driven Quantum Systems}

\author{Shintaro Takayoshi$^1$\thanks{takayoshi@konan-u.ac.jp}, 
and Takashi Oka$^2$}
\inst{
$^1$Department of Physics, Konan University, Kobe 658-8501, Japan \\
$^2$The Institute for Solid State Physics, The University of Tokyo, Kashiwa, Chiba 277-8581, Japan} %\\

\abst{We review quantum tunneling provoked by external field driving, 
focusing on the role of geometric effects. 
The discussion begins with an overview of tunneling phenomena, 
including the Landau-Zener model and the Schwinger effect, 
both of which are essential frameworks to describe 
the generation of elementary excitation of the system. 
We also refer to the relation between the modern theory of polarization 
and the geometry of the system,
and introduce the shift vector via adiabatic perturbation theory. 
Then we introduce the twisted Landau-Zener model and 
shown how the shift vector modulates tunneling probability, 
followed by several illustrative applications of this model. 
We also explain the Keldysh crossover, 
which is the crossover from a quantum tunneling regime to 
photon absorption regime in driven systems.}

\begin{document}
\maketitle

\section{Introduction}

Tunneling phenomena represent one of the oldest problems 
in quantum mechanics and have applications in various fields of physics. 
The Landau-Zener (LZ) model is a prototypical framework 
that describes transitions in two-level systems near a band 
anticrossing~\cite{Landau1932ZSow,Rosen1932PR,Stuckelberg1932,Zener1934ProcRoy}. 
In this model, as the rate of parameter change increases, 
nonadiabatic transitions become more prominent, 
and the tunneling probability rises monotonically. 
Since this problem is related with the manipulation of two-level states, 
it is important in the fields of spintronics~\cite{Miao2011RPP} 
and quantum computation~\cite{Gaitan2003PRA,Li2011JModOpt}. 
In quantum electrodynamics, a nonperturbative phenomenon of 
vacuum decay by the application of a strong electric field, 
which results in the spontaneous creation of 
particle-antiparticle pairs, has been proposed by 
Heisenberg and Euler~\cite{Heisenberg1936ZPhys} 
and later refined by Schwinger~\cite{Schwinger1951PR}. 
This phenomenon can be understood from the viewpoint of 
quantum tunneling in the Dirac model 
with the integration over the momentum space. 
This effect is important for understanding the interplay 
between quantum fluctuations and electromagnetic field, 
and its implications are applied to 
many significant quantum phenomena 
in high-energy physics, astrophysics, and condensed matter physics. 
such as dielectric breakdown in 
semiconductors~\cite{Zener1934ProcRoy,Kane1960JPCS} and
Mott insulators~\cite{Oka2003PRL,OkaAoki2005PRL,Oka2012PRB,Mayer2015PRB,Yamakawa2017NatMater,Takamura2023PRB}, 
and the Kibble-Zurek 
mechanism~\cite{Kibble1976JPhysA,Zurek1985Nature,Damski2005PRL}. 
The Schwinger effect can be extended to the AC 
fields~\cite{Brezen1970PRD,Popov1974SovJNP,SchutzholdPRL2008}, 
which is related with the problems of
strong field ionization~\cite{Ivanov2005JModOpt,Krausz2009RMP,Blaga2009NatPhys,Popruzhenko2014JPhysB} 
and a particle escaping from an oscillating trap~\cite{Keldysh1965JETP}.

Another topic gathering attention in recent years is 
the effect of geometry on physical properties. 
Geometric phases such as the 
Berry phase~\cite{Berry1984,Xiao2010RMP} play a crucial role 
in phenomena like the quantum Hall 
effect~\cite{Thouless1982PRL,Kohmoto1985AnnPhys} and 
topological insulators~\cite{Hasan2010RMP,Qi2011RMP}. 
Their influence on nonlinear electronic processes 
has also attracted interest, including the nonlinear 
Hall effect~\cite{Sodemann2015PRL} and 
the emergence of even-order nonlinear optical responses 
and shift currents in crystals with broken inversion 
symmetry~\cite{Sipe2000PRB,Morimoto2016SciAdv}. 
In addition, Floquet engineering~\cite{Oka2019AnnRev} 
provides a powerful framework for realizing topological states 
in nonequilibrium 
systems~\cite{Oka2009PRB,Kitagawa2011PRB,Lindner2011NatPhys}. 

The influence of geometric effects in dynamics goes beyond 
the contribution to the phase factor~\cite{Berry1984,Aharonov1987PRL}. 
It is pointed out that the geometry affects the tunneling probability 
in nonadiabatic transition processes through the geometric amplitude 
factor~\cite{Berry1990ProcRoy,Nakamura1994PRA,Bouwmeestert1996JModOpt}. 
This effect has been applied to the counterdiabatic driving of 
the system~\cite{Demirplak2003JPCA,Berry2009JPA,delCampo2013PRL}. 

%%%%%%%%%% Fig : Keldysh crossover %%%%%%%%%%
\begin{figure}[t]
\centering
\includegraphics[width=0.4\textwidth]{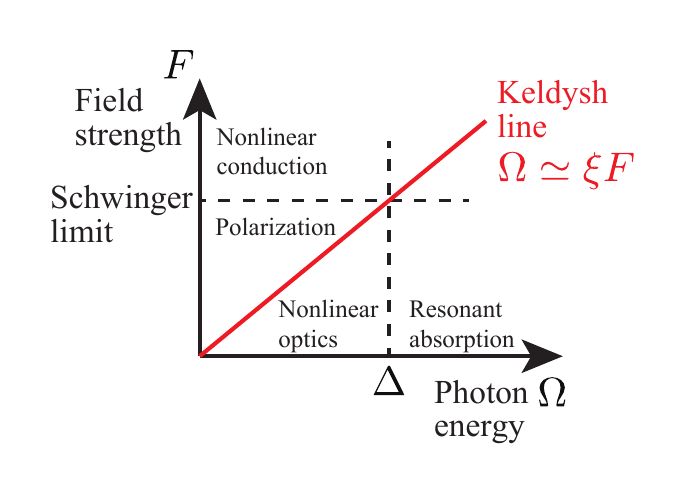}
\caption{Various phenomena provoked by the application of 
external AC fields in the parameter space of 
photon energy $\Omega$ and field strength $F$. 
$\xi$ and $\Delta$ are the correlation length and the gap 
of the system, respectively. 
The change of behavior over the Keldysh line 
$\Omega \simeq \xi F$ is called the Keldysh crossover.}
\label{fig:Keldysh}
\end{figure}
%%%%%%%%%%%%%%%%%%%%

When considering excitations of a system driven by an external field, 
two types of mechanisms come into play: 
quantum tunneling and photon absorption. 
The dominant mechanism is determined by 
the field strength $F$ and the angular frequency $\Omega$ 
of the driving field [Fig.~\ref{fig:Keldysh}]. 
The quantum tunneling regime and the photon absorption regime 
are separated by the Keldysh line $\Omega \simeq \xi F$, 
where $\xi$ represents the correlation length of the system. 
By varying these parameters, the transition between the two regimes, 
known as the Keldysh crossover, is observed~\cite{Keldysh1965JETP}. 

This paper aims to review quantum tunneling induced 
by external field driving and the role of geometry 
in nonadiabatic excitation, 
especially the geometric amplitude factor. 
To this end, we introduce a twisted LZ (TLZ) model, 
discuss its properties, and demonstrate several examples 
of its applications. 
We also address the Keldysh crossover in many-body systems.

This paper is organized as follows. 
In Sec.~\ref{sec:Geometry}, we review tunneling phenomena, 
particularly focusing on the LZ model and the Schwinger effect. 
We also discuss the modern theory of polarization, 
which is related to geometry, and introduce the concept of 
the shift vector by reviewing adiabatic perturbation theory. 
In Sec.~\ref{sec:TLZ}, we introduce the TLZ model and demonstrate 
how the shift vector influences the tunneling probability. 
Several applications of the TLZ model are presented 
in Sec.~\ref{sec:Application}. 
Section~\ref{sec:Crossover} provides an explanation 
of the Keldysh crossover. 
Finally, a summary and discussion is presented 
in Sec.~\ref{sec:Summary}.

\section{Excitation by tunneling phenomena and geometric effects}
\label{sec:Geometry}

\subsection{Quantum tunneling}

First, we explain about the excitations generated by 
quantum tunneling. 
One of the simplest models to describe 
the tunneling phenomena is the LZ 
model~\cite{Landau1932ZSow,Zener1934ProcRoy,Stuckelberg1932}: 
\begin{align}
 \hat{\mathcal{H}}_{\mathrm{LZ}}(t)
   =m\hat{\sigma}^{z}+vq(t)\hat{\sigma}^{x},\quad
 q(t)=-Ft,
\label{eq:LZmodel}
\end{align}
where $m$ and $v$ are constant, 
$\hat{\sigma}^{j}$ is the Pauli matirices, 
and $q(t)$ is swept linearly 
in the time interval $-\infty<t<\infty$. 
In this paper, we employ the unit $c=\hbar=1$. 
The profile of the instantaneous Hamiltonian is 
a two-level system with avoided crossing, 
and the gap is $\sqrt{m^{2}+(vq(t))^{2}}$ taking the minimum at $t=0$. 
During this process, the lower energy initial state 
makes a tunneling transition to the higher energy one, 
and its probability is obtained analytically as 
\begin{align}
 P(F)=\exp\Big[-\pi \frac{m^{2}}{|vF|}\Big].
\end{align}
In the limit of $|F| \to 0$, $P(F)$ approaches zero, 
which is the adiabatic limit.

In the quantum electrodynamics, an analogous phenomenon 
happens in the creation process of electron-positron pairs from the vacuum
in the strong DC electric field $\boldsymbol{F}$, 
so-called the Schwinger effect. 
It can be formulated as a massive Dirac model 
\begin{align}
 \hat{\mathcal{H}}_{\mathrm{D}}
   =v\sum_{j=x,y,z}\hat{\gamma}^{0}\hat{\gamma}^{j}(k_{j}+e_{0}A_{j})
     +m\hat{\gamma}^{0},
\label{eq:Dirac3D}
\end{align}
where $-e_{0}$ is the charge of electron, 
$\boldsymbol{A}=-\boldsymbol{F}t$ is the vector potential, 
and $\hat{\gamma}^{\mu}$ $(\mu=0,x,y,z)$ are the gamma matrices 
$\hat{\gamma}^{0}=
\begin{pmatrix}
  0& I  \\
  I& 0
\end{pmatrix}$, 
$\hat{\gamma}^{j}=
\begin{pmatrix}
  0& \hat{\sigma}^{j}  \\
  -\hat{\sigma}^{j}& 0
\end{pmatrix}$ 
($j=x,y,z$). 
For each wave number $\boldsymbol{k}$, 
the generation of an electron-positron pair is regarded 
as a tunneling process from the lower to upper band. 
Schwinger estimated the total pair production rate 
at the one-loop order as~\cite{Schwinger1951PR} 
\begin{align}
 \Gamma \propto \sum_{n=1}^{\infty}\frac{1}{n^{2}}
   \exp\Big(-\frac{\pi m^{2}n}{ve_{0}|\boldsymbol{F}|}\Big).
\label{eq:Schwin}
\end{align}
Thus electron-positron pairs proliferate 
when the electric field strength is above the threshold 
\begin{align}
 F_{\mathrm{th}}=m^{2}/(ve_{0}),
\label{eq:Schwinlim}
\end{align}
which is called the Schwinger 
limit~\cite{Sauter1931ZPhys,Heisenberg1936ZPhys,Schwinger1951PR}.

\subsection{Schwinger effects and polarization from the geometric viewpoint}

Next we explain the modern formulation of polarization 
and discuss its relation with the Schwinger effect. 
King-Smith and Vanderbilt proposed a expression of polarization 
in terms of a geometric property of 
states~\cite{KingSmith1993PRB,Vanderbilt1993PRB}: 
\begin{align}
 \boldsymbol{P}_{\mathrm{e}}
   =-\frac{e_{0}}{(2\pi)^3} \sum_{n:\mathrm{occ}} \int_{\mathrm{BZ}} 
     i\braket{u_{n,\boldsymbol{k}}|\nabla_{\boldsymbol{k}}
             |u_{n,\boldsymbol{k}}}
d\boldsymbol{k},
\end{align}
where 
$i\braket{u_{n,\boldsymbol{k}}|\nabla_{\boldsymbol{k}}|u_{n,\boldsymbol{k}}}$ 
represents the Berry connection, 
$u_{n,\boldsymbol{k}}$ is the periodic part of the wave function, 
and the summation is taken over the occupied bands. 
In the following, we focus on the 1D case for simplicity, where 
$P_{\mathrm{e}}=-\frac{e_{0}}{2\pi} \sum_{n:\mathrm{occ}}
\int_{\mathrm{BZ}} i\braket{u_{n,k}|\partial_{k}|u_{n,k}}dk$. 
This expression avoids ambiguities in crystals under 
the periodic boundary condition 
in contrast to the traditional way of calculation 
based on charge distributions. 
Resta extended this approach to general quantum 
states~\cite{Resta1998PRL,Resta1999PRL}
using the twist operator: 
\begin{align}
 P_{\mathrm{e}} = \frac{e_{0}}{2\pi}
   \mathrm{Im} \ln 
   \braket{\Psi_{0}| e^{i\frac{2\pi}{L} \hat{X}} |\Psi_{0}},
\label{eq:PolResta}
\end{align} 
where $\hat{X}=\sum_{r}r\hat{n}_{r}$ 
is the position operator 
($\hat{n}_{r}$ is the number operator) and 
$L$ is the system size. 
The twist operator $e^{i\frac{2\pi}{L} \hat{X}}$ 
shifts the state in the momentum space, 
and it is used for discussing the Lieb-Schultz-Mattis 
theorem~\cite{Lieb1961AnnPhys,Oshikawa1997PRL,Oshikawa2000PRL,Hastings2005Europhys}. 
The polarization of one-dimensional (1D) insulators acts as a topological index 
under the symmetry constraint of 
the system~\cite{Nakamura2002PRB,Nakamura2002PRL}.
For example, if the system has inversion symmetry, 
the polarization value takes discrete values of $0$ or $1/2$, 
which indicate whether the system is in a trivial or topological phase.

Now we discuss the relation between Schwinger effects 
and polarization in terms of the geometry~\cite{OkaAoki2005PRL,OkaAoki2010PRB}. 
The Hamiltonian of a system in a DC electric field 
in the potential gauge is given as
$\hat{\mathcal{H}}=\hat{\mathcal{H}}_{0}+e_{0}F\hat{X}$. 
We consider the time evolution by the Hamiltonian $\hat{\mathcal{H}}$ 
with the initial state $\ket{\Psi_{0}}$, 
which is the ground state of $\hat{\mathcal{H}}_{0}$. 
Then the state is given by 
$\ket{\Psi(t)}=e^{-it(\hat{\mathcal{H}}_{0}+e_{0}F\hat{X})}\ket{\Psi_{0}}$. 
The transition amplitude 
from $\ket{\Psi_{0}}$ to $e^{-i\hat{\mathcal{H}}_{0}t}\ket{\Psi_{0}}$ 
through this time evolution is represented as 
\begin{align}
 \Xi(t)=&\braket{\Psi_{0}|e^{it\hat{\mathcal{H}}_{0}}|\Psi(t)}
\nonumber\\
   =&\braket{\Psi_{0}|e^{-it(\hat{\mathcal{H}}_{0}+e_{0}F\hat{X})}
           |\Psi_{0}} e^{itE_{0}},
\label{eq:XiBraket}
\end{align}
where $E_{0}$ is the ground state energy of 
$\hat{\mathcal{H}}_{0}$. 
The effective Lagrangian can be asymptotically  defined as 
\begin{align}
 \Xi(t) \sim e^{itL\mathcal{L}(F)}.
\label{eq:XiLagrangian}
\end{align}
In Eq.~\eqref{eq:XiBraket}, 
\begin{align}
 \braket{\Psi_{0}|e^{-it(\hat{\mathcal{H}}_{0}+e_{0}F\hat{X})}|\Psi_{0}}
   =\braket{\Psi_{0}|e^{-it\hat{\mathcal{H}}_{0}}e^{-ite_{0}F\hat{X}}
     e^{\mathcal{O}(F^{2})}|\Psi_{0}},
\end{align}
from the Baker-Campbell-Hausdorff formula 
since $\hat{\mathcal{H}}_{0}$ is just a c-number $E_{0}$ 
for the commutation relation containing only one $e_{0}F\hat{X}$ term
inside the expectation value $\braket{\Psi_{0}|\cdot|\Psi_{0}}$. 
Here we set $t=2\pi/(Le_{0}F)$ and take the limit of $F \to 0$, then 
$\lim_{F \to 0}\Xi(2\pi/(Le_{0}F))
=\braket{\Psi_{0}|e^{-i\frac{2\pi}{L}\hat{X}}|\Psi_{0}}
=(\braket{\Psi_{0}|e^{i\frac{2\pi}{L}\hat{X}}|\Psi_{0}})^{*}$. 
Hence, the transition probability is related with 
the polarization Eq.~\eqref{eq:PolResta} as 
\begin{align}
 P_{\mathrm{e}} = -\lim_{F \to 0}\frac{e_{0}}{2\pi}
   \mathrm{Im} \ln \Xi\Big(\frac{2\pi}{Le_{0}F}\Big).
\end{align}
By using the effective Lagrangian Eq.~\eqref{eq:XiLagrangian}, 
this expression is recast into 
\begin{align}
 P_{\mathrm{e}} 
   =-\lim_{F \to 0}\frac{1}{F} \mathrm{Re} \mathcal{L}(F)
   =-\frac{\partial \mathrm{Re} \mathcal{L}(F)}{\partial F}
     \bigg|_{F=0}.
\end{align}

\subsection{Geometric effects in the adiabatic perturbation theory}

In this subsection, we describe the Geometric effects 
from the viewpoint of the adiabatic perturbation theory. 
We consider the time-dependent Hamiltonian of two-level systems
\begin{align}
 \hat{\mathcal{H}}(t)
   =\boldsymbol{d}(q(t)) \cdot \hat{\boldsymbol{\sigma}},
\label{eq:TwoLevelHamil}
\end{align}
where $q(t)=k+e_{0}Ft$. 
It is assumed that the gap does not close for the whole time interval. 
The instantaneous eigenenergy $E_{\pm}(t)$ ($E_{+}(t)>E_{-}(t)$) 
and eigenstates $\ket{u_{\pm}(t)}$ are defined as 
\begin{align}
 \hat{\mathcal{H}}(t)\ket{u_{\pm}(t)}
   =E_{\pm}(t)\ket{u_{\pm}(t)}.
\end{align}
We expand the solution of the Schr\"odinger equation 
in the basis of $\ket{u_{\pm}(t)}$ as 
\begin{align}
 \ket{\psi(t)}
   =\sum_{n=\pm}a_{n}(t)e^{-i\int_{0}^{t}ds\{E_{n}(s)-e_{0}FA_{nn}(s)\}}
     \ket{u_{n}(t)},
\label{eq:SolExpandBP}
\end{align}
where 
\begin{align}
 A_{nm}(t)=\braket{u_{n}(t)|i\partial_{q}|u_{m}(t)}
   =\frac{1}{e_{0}F}\braket{u_{n}(t)|i\partial_{t}|u_{m}(t)}
\end{align}
stands for the Berry connection. 
Substituting Eq.~\eqref{eq:SolExpandBP} into the Schr\"odinger equation 
\begin{align}
 i\partial_{t}\ket{\psi(t)}=\hat{\mathcal{H}}(t)\ket{\psi(t)},
\label{eq:SchEq}
\end{align}
we obtain 
\begin{align}
 &\sum_{n=\pm}(i\partial_{t}a_{n}(t))
   e^{-i\int_{0}^{t}ds\{E_{n}(s)-e_{0}FA_{nn}(s)\}}
   \ket{u_{n}(t)}
\nonumber\\
 &-\sum_{n=\pm}a_{n}(t)e^{-i\int_{0}^{t}ds\{E_{n}(s)-e_{0}FA_{nn}(s)\}}
   e_{0}FA_{nn}(t)\ket{u_{n}(t)}
\nonumber\\
 &+\sum_{n=\pm}a_{n}(t)e^{-i\int_{0}^{t}ds\{E_{n}(s)-e_{0}FA_{nn}(s)\}}
   (i\partial_{t}\ket{u_{n}(t)})=0.
\label{eq:EqCoeff}
\end{align}
By applying $\bra{u_{+}(t)}$ and $\bra{u_{-}(t)}$
to Eq.~\eqref{eq:EqCoeff}, 
we derive the equations 
\begin{align}
 &i\partial_{t}a_{+}(t)=-a_{-}(t)e_{0}FA_{+-}(t)
\nonumber\\
   &\quad \times
     e^{i\int_{0}^{t}ds\{E_{+}(s)-E_{-}(s)
     -e_{0}FA_{++}(s)+e_{0}FA_{--}(s)\}},\\
 &i\partial_{t}a_{-}(t)=-a_{+}(t)e_{0}FA_{-+}(t)
\nonumber\\
   &\quad \times
     e^{-i\int_{0}^{t}ds[E_{+}(s)-E_{-}(s)
     -e_{0}FA_{++}(s)+e_{0}FA_{--}(s)]}.
\end{align}
Here we introduce a gauge-independent quantity 
called geometric amplitude factor
\begin{align}
 R_{nm}(t)=-A_{nn}(t)+A_{mm}(t)+\partial_{q}\arg A_{nm}(t),
\label{eq:GeoAmpFac}
\end{align}
which is introduced by Berry~\cite{Berry1990ProcRoy}. 
Then the equations become 
\begin{align}
 &i\partial_{t}a_{+}(t)
   =-a_{-}(t)e_{0}F|A_{+-}(t)|
\nonumber\\
     &\quad\times e^{i\int_{0}^{t}ds\{E_{+}(s)-E_{-}(s)
     +e_{0}FR_{+-}(s)\}+i\arg A_{+-}(0)},
\\
 &i\partial_{t}a_{-}(t)
   =-a_{+}(t)e_{0}F|A_{-+}(t)|
\nonumber\\
     &\quad\times e^{-i\int_{0}^{t}ds\{E_{+}(s)-E_{-}(s)
     -e_{0}FR_{-+}(s)\}+i\arg A_{-+}(0)}.
\end{align}
Here we have used 
$e^{-i\arg A_{+-}(t)}A_{+-}(t)=|A_{+-}(t)|$ and 
$e^{-i\arg A_{-+}(t)}A_{-+}(t)=|A_{-+}(t)|$. 
With the initial condition $a_{-}(0)=1$ and $a_{+}(0)=0$, 
the solution is written in the integral form as 
\begin{align}
 &a_{+}(t)
   =i\int_{0}^{t}ds
     a_{-}(s)e_{0}F|A_{+-}(s)|
\nonumber\\
     &\;\times e^{i\int_{0}^{s}ds'\{E_{+}(s')-E_{-}(s')
     +e_{0}FR_{+-}(s')\}+i\arg A_{+-}(0)},
\\
 &a_{-}(t)
   =i\int_{0}^{t}ds
     a_{+}(s)e_{0}F|A_{-+}(s)|
\nonumber\\
     &\;\times e^{-i\int_{0}^{s}ds'\{E_{+}(s')-E_{-}(s')
     -e_{0}FR_{-+}(s')\}+i\arg A_{-+}(0)}.
\end{align}
The formal solution is given by 
substituting $a_{\pm}$ in the right hand sides iteratively. 
Using the first order iterative approximation, 
i.e., $a_{-}(t)\simeq 1$, 
\begin{align}
 a_{+}(t)
   &=ie^{i\arg A_{+-}(0)}\int_{0}^{t}ds e_{0}F|A_{+-}(s)|
\nonumber\\
 &\times e^{i\int_{0}^{s}ds' \{E_{+}(s')-E_{-}(s')+e_{0}FR_{+-}(s')\}}.
\label{eq:aplus}
\end{align}
The integration in Eq.~\eqref{eq:aplus} can be 
evaluated by the Dykhne-Davis-Pechukas (DDP) 
method~\cite{Dykhne1962JETP,Davis1976JCP,Joye1991AnnPhys} 
also known as the Landau-Dykhne method. 
In the DDP method, 
the integration is performed in the space of wave number 
extended to the complex plane. 
In this complex wave number space, 
an exceptional point $\boldsymbol{k}_{\mathrm{c}}$, 
where the excitation gap is closed, 
can be reached by proceeding in the imaginary direction 
from the gap minimum point on the real axis $\boldsymbol{k}_{0}$. 
Recently the extension of the DDP formula is discussed 
in terms of the Lefschetz-thimble analysis~\cite{Fukushima2020AnnPhys}.
After evaluating the integration in Eq.~\eqref{eq:aplus} by the DDP method, 
the tunneling probability is calculated to be 
\begin{align}
 P(F)&=|a_{+}(t)|^{2}
\nonumber\\
   &\simeq \exp\Big[2\mathrm{Im}
     \int_{\boldsymbol{k}_{0}}^{\boldsymbol{k}_{\mathrm{c}}}
     d\boldsymbol{k}
     \frac{E_{+}-E_{-}+e_{0}FR_{+-}}{e_{0}|F|}
     \Big].
\label{eq:ProbR}
\end{align}
This formula indicates that the tunneling probability 
is modulated by the geometric amplitude factor 
as pointed out by Berry~\cite{Berry1990ProcRoy}. 
The expression of $E_{+}-E_{-}$, $|A_{+-}|$, and $R_{+-}$ 
in Eq.~\eqref{eq:TwoLevelHamil} 
for the parameter curve $\boldsymbol{d}(q(t))$ 
is provided in Ref.~\cite{Kitamura2020CommPhys}. 
We give the extension of Eq.~\eqref{eq:ProbR} 
which includes the $\mathcal{O}(R_{+-}^{2})$ term 
in Sec.~\ref{sec:TLZ}.

Geometric amplitude factor Eq.~\eqref{eq:GeoAmpFac} 
is also called quantum geometric 
potential~\cite{Wu2008PRA,Xu2018PRA}. 
In the expression of polarization in terms of the Berry phase, 
$R_{nm}(q)$, which is Eq.~\eqref{eq:GeoAmpFac} represented 
as a function of the crystal momentum $q$, 
is known as the shift vector. 
The shift vector $R_{+-}$ quantifies the difference of 
the electric polarization 
between the upper and lower bands~\cite{Sipe2000PRB}. 
In Ref.~\cite{Kitamura2020CommPhys}, 
Eq.~\eqref{eq:ProbR} is applied to the Rice-Mele model 
to discuss the nonreciprocal tunneling. 
The multiple tunneling effect in nonreciprocal LZ tunneling 
has been studied recently~\cite{Terada2025PRB}.

\section{Twisted LZ model}
\label{sec:TLZ}

To demonstrate the nonadiabatic geometric effects, 
we introduce a simple time-dependent Hamiltonian 
of a two-level system, 
\begin{align}
 \hat{\mathcal{H}}(q(t))
   =m\hat{\sigma}^{z}+vq(t)\hat{\sigma}^{x}
     +\frac{1}{2}\kappa_{\parallel} v^{2}q(t)^{2}\hat{\sigma}^{y},
\quad q(t)=-Ft
\label{eq:Hamil_q2}
\end{align}
where $m$ and $v$ are constant. 
Since the $q(t)^{2}$ term is added to the LZ model 
Eq.~\eqref{eq:LZmodel}, we call Eq.~\eqref{eq:Hamil_q2} 
the TLZ model. 

In the notation of Eq.~\eqref{eq:TwoLevelHamil}, 
\begin{align}
 \boldsymbol{d}(q(t))
   =(vq(t), \frac{1}{2}\kappa_{\parallel} v^{2}q(t)^{2}, m)
\end{align}
draws a parabolic trajectory in the three-dimensional (3D) space 
while the standard LZ model ($\kappa_{\parallel}=0$) 
draws a straight line. 
Note that $\kappa_{\parallel}$ is the curvature of the parabola 
at the vertex, where the gap of the system takes the minimum.

\subsection{Tunneling probability}

We analyze the tunneling probability of Eq.~\eqref{eq:Hamil_q2}, 
by utilizing the time-dependent unitary transformation, 
which corresponds to changing the frame of the system. 
Let us consider the unitary transformation 
$\ket{\psi(t)}=\hat{U}(t)\ket{\psi'(t)}$. 
Substituting this relation into 
the Schr\"odinger equation Eq.~\eqref{eq:SchEq}, 
we obtain 
$i\partial_{t}\hat{U}(t))\ket{\psi'(t)}+\hat{U}(t)i\partial_{t}\ket{\psi'(t)}
=\hat{\mathcal{H}}(t)\hat{U}(t)\ket{\psi'(t)}$, 
which is finally rewritten as 
\begin{align}
 i\partial_{t}\ket{\psi'(t)}
   =\big[\hat{U}^{\dagger}(t)\hat{\mathcal{H}}(t)\hat{U}(t)
     -\hat{U}^{\dagger}(t)(i\partial_{t}\hat{U}(t))\big]
     \ket{\psi'(t)}.
\end{align}
Hence the effective Hamiltonian in the $\ket{\psi'(t)}$ frame is 
\begin{align}
 \hat{\mathcal{H}}_{\mathrm{eff}}(t)
   =\hat{U}^{\dagger}(t)\hat{\mathcal{H}}(t)\hat{U}(t)
     -\hat{U}^{\dagger}(t)(i\partial_{t}\hat{U}(t)).
\label{eq:RotHeff}
\end{align}
This time-dependent unitary transformation is useful 
for analyzing the dynamics of quantum systems 
in terms of an effective static 
model~\cite{Takayoshi2014PRBa,Takayoshi2014PRBb}.

\begin{figure}[t]
\centering
\includegraphics[width=0.48\textwidth]{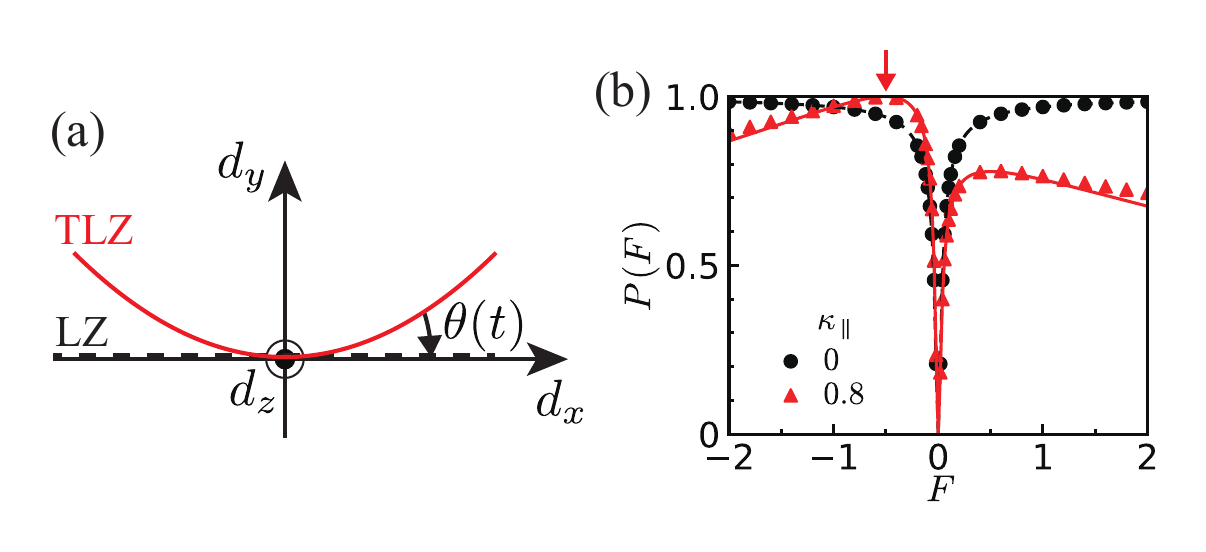}
\caption{(a) Time-dependent rotation which transforms 
the TLZ model into the LZ model. 
(b) Tunneling probability 
of the model Eq.~\eqref{eq:Hamil_q2} with $m=0.1$ and $v=1$. 
Numerically calculated $P(F)$ is shown by 
circles (the LZ model $\kappa_{\parallel}=0$) and 
triangles (the TLZ model $\kappa_{\parallel}=0.8$).
The dashed and solid lines are the predictions from 
Eq.~\eqref{eq:LZquad} for the LZ and TLZ models, respectively.
The arrow represents the sweeping speed 
at which the prefect tunneling happens Eq.~\eqref{eq:PTspeed}.}
\label{fig:TLZ}
\end{figure}

In the present case, we transform the TLZ model 
into the standard LZ model 
using a time-dependent rotation about the $d_{z}$ axis 
$\hat{U}(t)=e^{i\frac{\theta(q(t))}{2}\hat{\sigma}^{z}}$, 
where 
\begin{align}
 \theta(q(t))=-\arctan\frac{d_{y}(q(t))}{d_{x}(q(t))}
   =-\arctan\frac{1}{2}\kappa_{\parallel} vq(t)
\label{eq:thetat}
\end{align}
is the rotation angle 
[Fig.~\ref{fig:TLZ}(a)]. 
Since the tunneling happens mainly around 
the gap minimum point $q=0$, we expand 
\begin{align}
 \theta(q)\simeq
   -\frac{1}{2}\kappa_{\parallel} vq+\mathcal{O}(q^{3}).
\end{align}
Substituting Eq.~\eqref{eq:thetat} into Eq.~\eqref{eq:RotHeff}, 
we derive an effective Hamiltonian 
\begin{align}
 \hat{\mathcal{H}}_{\mathrm{eff}}(t)
   =\Big(m+\frac{\kappa_{\parallel}vF}{4}\Big)
     \hat{\sigma}^{z}
     +\big(vq+\mathcal{O}(q^{3})\big)\hat{\sigma}^{x},
\end{align}
where the $\hat{U}^{\dagger}(t)(i\partial_{t}\hat{U}(t))$ term gives 
$-\frac{1}{2}\frac{dq}{dt}\frac{d\theta(q)}{dq}\hat{\sigma}^{z}
=(-\frac{\kappa_{\parallel}vF}{4}+\mathcal{O}(q^{3}))\hat{\sigma}^{z}$. 
Therefore, for a linear sweep $q=-Ft$, 
the tunneling probability $P(F)$ 
in Eq.~\eqref{eq:Hamil_q2} becomes 
\begin{align}
 P(F)=\exp\bigg[
   -\pi
     \frac{(m+\kappa_{\parallel}vF/4)^{2}}{|vF|}\bigg].
\label{eq:LZquad}
\end{align}
For the model Eq.~\eqref{eq:Hamil_q2}, 
the geometric amplitude factor Eq.~\eqref{eq:GeoAmpFac} 
is calculated to be $R_{+-}(t=0)=\kappa_{\parallel}v$. 
Hence, Eq.~\eqref{eq:LZquad} is rewritten as 
\begin{align}
 P(F)=\exp\bigg[
   -\pi \frac{(m+R_{+-}F/4)^{2}}{|vF|}\bigg].
\label{eq:PF_TLZ}
\end{align}
In Eq.~\eqref{eq:PF_TLZ}, 
we notice that the tunneling gap is effectively modified 
by the geometric amplitude factor as 
$\Delta_{\mathrm{eff}}=2m+\kappa_{\parallel}vF/2$. 
This result is the improvement of Eq.~\eqref{eq:ProbR} with $e_{0}=1$ 
by including the $\mathcal{O}(R_{+-}^{2})$ term in the exponential. 

%%%%%%%%%% Fig : Tunneling probability %%%%%%%%%%
%%%%%%%%%%%%%%%%%%%%

Here, we have considered the case that 
the positional vector of the gap minimum in the parameter space 
$(0,0,m)$ is perpendicular to the plane which contains 
the neighborhood of the curve $\boldsymbol{d}(q)$ 
around the gap minimum. 
We can extend the above discussion to the general parameter curve 
\begin{align}
 \hat{\mathcal{H}}(t)=\hat{A}+\hat{B}q(t)+\hat{C}q(t)^{2}/2,
\label{eq:GeneralCurve}
\end{align}
where the conditions $\{\hat{A},\hat{B}\}=0$ and 
$\{\hat{B},\hat{C}\}=0$ are required in order for 
the energy gap and the absolute value of velocity to be minimal 
at $q(t)=0$. We can apply Eq.~\eqref{eq:LZquad} 
to the curve Eq.~\eqref{eq:GeneralCurve} by using 
the parameters~\cite{Takayoshi2021SciPost} 
\begin{align}
 m=\|\hat{A}\|,\quad
 v=\|\hat{B}\|,\quad
 \kappa_{\parallel}v^{2}=-\frac{i}{8}
   \frac{\mathrm{Tr}\{[\hat{A},\hat{B}],\hat{C}\}}{\|\hat{A}\|\|\hat{B}\|},
\label{eq:TLZgeneral}
\end{align}
where 
$\|\hat{O}\|\equiv\frac{1}{2}\sqrt{\mathrm{Tr}\{\hat{O},\hat{O}\}}$.

\subsection{Numerical demonstration}

We numerically calculate the time evolution of 
the TLZ model Eq.~\eqref{eq:Hamil_q2} 
and evaluate the tunneling probability 
for various sweeping speed. 
In Fig.~\ref{fig:TLZ}(b), 
the circles and triangles show the numerically calculated results, 
and the dashed and solid lines show the predictions from 
Eq.~\eqref{eq:LZquad}. 
Comparing the circles and dashed line 
(the LZ model $\kappa_{\parallel}=0$) 
with the triangles and solid line 
(the TLZ model $\kappa_{\parallel}=0.8$), 
we notice several anomalous phenomena in the TLZ model. 

The first is the rectification or nonreciprocity, 
i.e., the tunneling probability depends on 
the direction of the sweep. 
From Eq.~\eqref{eq:LZquad}, 
the ratio of tunneling probabilities between two sweep directions 
is evaluated as 
\begin{align}
 P(|F|)/P(-|F|)=\exp(-\pi m \kappa_{\parallel}).
\end{align}
This phenomenon arises from the $\mathcal{O}(\kappa_{\parallel})$ term 
in the exponential and has been indicated 
by Berry~\cite{Berry1990ProcRoy}. 

The second is that the perfect tunneling $P=1$ takes place 
as shown by an arrow in Fig.~\ref{fig:TLZ}(b). 
As we can see from Eq.~\eqref{eq:LZquad}, 
it happens when the sweeping speed is 
\begin{align}
 F_{\mathrm{PT}}
   =-\frac{4m}{\kappa_{\parallel}v},
\label{eq:PTspeed}
\end{align}
where the modified gap 
$\Delta_{\mathrm{eff}}=2m+\kappa_{\parallel}vF/2$ 
becomes zero and the system effectively shows a gapless behavior. 

The third is the counterdiabaticity. 
In Fig.~\ref{fig:TLZ}(b), 
$P(F)$ decreases as $|F|$ increases in the large $|F|$ region 
for the TLZ case 
while $P(F)$ increases monotonically as $|F|$ increases 
for the standard LZ case. 
From Eq.~\eqref{eq:LZquad}, 
the tunneling probability behaves as 
$P(F)\simeq \exp(-\pi\kappa_{\parallel}^{2}|vF|/16)$ 
in the large $|F|$ regime.

\section{Application of twisted LZ model}
\label{sec:Application}

\subsection{Single spin in time-dependent external magnetic field}

The simplest application of the TLZ model 
is the isolated spin-1/2 in the magnetic field 
$\boldsymbol{h}(t)$. 
The external field is static in the $z$ direction 
and time-dependent on the $xy$ plane: 
$\boldsymbol{h}(t)=(\alpha t, \beta t^{2}, h_{z})$. 
The Hamiltonian is given as 
\begin{align}
 \hat{\mathcal{H}}(t)
   =-\alpha t \hat{S}^{x}-\beta t^{2}\hat{S}^{y}
     -h_{z}\hat{S}^{z},
\label{eq:spinTLZ}
\end{align}
which is the same as the TLZ model Eq.~\eqref{eq:Hamil_q2}.
For simplicity, we assume 
$\alpha>0$, $\beta>0$, and $h_{z}>0$. 
At the initial $t \to -\infty$ and final $t \to \infty$ time, 
the lower (higher) energy state corresponds to the spin 
in the $+y$ ($-y$) direction. 
Thus, through the TLZ process, 
the tunneling from the lower to higher energy state 
is the spin-flip. 
Finding an optimal modulation of external field is 
an important task for the high-speed spin control. 
From such viewpoint, counterdiabatic driving of quantum states 
have been studied~\cite{Torrontegui2013AdvOpt,Odelin2019RMP}. 
The magnetization process in quantum spin systems is also regarded as 
tunneling~\cite{Miyashita1995JPSJ,Raedt1997PRB}. 

Recently, spin control by the TLZ process is realized 
in nitrogen-vacancy centers of diamond~\cite{Sasaki2023PRA}. 
When an electron is captured at a point defect consisting of 
a nitrogen atom and a vacancy in diamond, 
the spin triplet states $m_{\mathrm{S}}=1,0,-1$ are formed. 
They use $m_{\mathrm{S}}=0$ and $-1$ states as a two-level system 
and realize the TLZ model Eq.~\eqref{eq:spinTLZ} 
by applying an appropriate microwave pulse. 
The tunneling probability is measured by changing 
the sweep velocity and the results show the behavior 
of the TLZ model including the nonreciprocity and 
the perfect tunneling.

\subsection{Saw-tooth chain model}

%%%%%%%%%% Fig : Tunneling probability %%%%%%%%%%
\begin{figure}[t]
\centering
\includegraphics[width=0.4\textwidth]{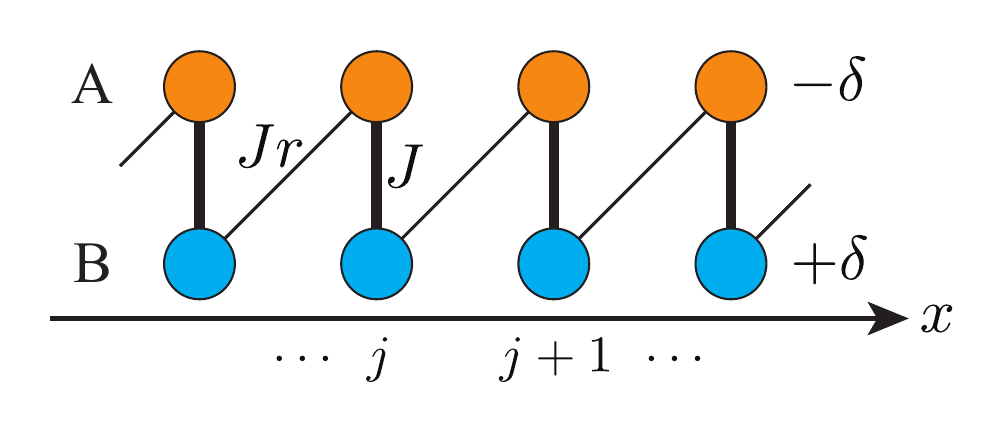}
\caption{The schematic picture of the saw-tooth chain 
model Eq.~\eqref{eq:HSaw}.}
\label{fig:SawTooth}
\end{figure}
%%%%%%%%%%%%%%%%%%%%

Next we consider a simple 1D lattice model 
which can be understood in terms of TLZ mechanism. 
As shown in Fig.~\ref{fig:SawTooth}, 
the lattice is a two-leg ladder with the rung index $j$ and 
the chain index $\{\mathrm{A,B}\}$. 
The hopping only exists between the sites 
$(j,\mathrm{A})$ and $(j,\mathrm{B})$,
and between $(j,\mathrm{B})$ and $(j+1,\mathrm{A})$ 
like a shape of the saw-tooth, 
thus we call this model as the saw-tooth chain model. 
The Hamiltonian is represented as
\begin{align}
 \hat{\mathcal{H}}=\sum_{j}
   [&-J(\hat{a}_{j,\mathrm{A}}^{\dagger}\hat{a}_{j,\mathrm{B}} +\mathrm{H.c.})
    -Jr(\hat{a}_{j,\mathrm{B}}^{\dagger}\hat{a}_{j+1,\mathrm{A}}+\mathrm{H.c.})
\nonumber\\
    &-\delta \hat{a}_{j,\mathrm{A}}^{\dagger}\hat{a}_{j,\mathrm{A}}
    +\delta \hat{a}_{j,\mathrm{B}}^{\dagger}\hat{a}_{j,\mathrm{B}}].
\label{eq:HSaw}
\end{align}
After the Fourier transformation 
\begin{align}
 \hat{a}_{k}=\sqrt{\frac{1}{N}}
   \sum_{j}e^{ikj}\hat{a}_{j,\mathrm{A}},\quad
 \hat{b}_{k}=\sqrt{\frac{1}{N}}
   \sum_{j}e^{ikj}\hat{a}_{j,\mathrm{B}},
\end{align}
the Hamiltonian is written in the form of 
\begin{align}
 \mathcal{H}=&\sum_{k}
\begin{pmatrix}
 \hat{a}_{k}^{\dagger} & \hat{b}_{k}^{\dagger}
\end{pmatrix}
\hat{\mathcal{H}}(k)
\begin{pmatrix}
 \hat{a}_{k} \\  \hat{b}_{k}
\end{pmatrix}
,
\end{align}
where
\begin{align}
 \hat{\mathcal{H}}(k)=&
\begin{pmatrix}
 -\delta & -J(1+re^{ ik}) \\
 -J(1+re^{-ik}) & \delta
\end{pmatrix}
\nonumber\\
 =&-J(1+r\cos k)\hat{\sigma}^{x}+(Jr\sin k)\hat{\sigma}^{y}
   -\delta\hat{\sigma}^{z}.
\label{eq:Hamil_SawTooth}
\end{align}
When the uniform static electric field $F$ is applied 
to this system, the Hamiltonian becomes 
$\hat{\mathcal{H}}(k+A)=\hat{\mathcal{H}}(k-Ft)$ 
(we set $e_{0}=1$). 
We expand Eq.~\eqref{eq:Hamil_SawTooth} 
around the gap minimum $k=\pi$ up to the second order 
\begin{align}
 \mathcal{H}(k')
   =-J(1-r)\hat{\sigma}^{x}-\delta\hat{\sigma}^{z}
     -Jr\hat{\sigma}^{y}k'
     -\frac{Jr}{2}\hat{\sigma}^{x}{k'}^{2},
\end{align}
where $k'=k-\pi$. 
Thus the TLZ model is realized for each crystal wavenumber $k$ 
and we can evaluate the tunneling probability $\mathcal{P}(F,k)$. 
In the actual model, the gap minimum is located periodically 
$k=\pi(2n+1)$ ($n$: integer) and the interband interference happens. 
Furthermore, the population of electrons in the upper and lower bands 
is determined from the relaxation process. 
Here, for simplicity, we assume that the population of 
electrons in the upper band
is proportional to $\mathcal{P}(F,k)$, 
which corresponds to the application of 
the relaxation time approximation. 
The improvement of the relaxation time approximation 
in the LZ tunneling has been proposed 
recently~\cite{Terada2024PRB,Terada2025pss}. 
Since the physical quantities are calculated through the integration over 
the crystal wave number $k$, 
we can observe  the ``twisted Schwinger effect'' in this system. 
For example, electric current is proportional to 
\begin{align}
 J_{\mathrm{c}}(F) \propto \int dk k\mathcal{P}(F,k).
\end{align}
We substitute
\begin{align}
 &m=\sqrt{J^{2}(1-r)^{2}+\delta^{2}},\quad
 v=Jr\nonumber\\
 &\kappa_{\parallel}v^{2}
   =\frac{Jr\delta}{\sqrt{J^{2}(1-r)^{2}+\delta^{2}}}
\end{align}
which is obtained from Eq.~\eqref{eq:TLZgeneral}, 
into Eq.~\eqref{eq:LZquad}. 
The tunneling probability becomes 
\begin{align}
 \mathcal{P}(F,k)
   =\exp\bigg[ & -\frac{\pi}{Jr|F|}
   \bigg\{\sqrt{J^{2}(1-r)^{2}+\delta^{2}}
\nonumber\\
     &-\frac{F\delta}{4\sqrt{J^{2}(1-r)^{2}+\delta^{2}}}\bigg\}^{2}\bigg],
\end{align}
which does not depend on $k$. 
Therefore the rectification is predicted to happen 
in this system as 
\begin{align}
 \frac{J_{\mathrm{c}}(|F|)}{J_{\mathrm{c}}(-|F|)}
   =\exp\Big(\frac{\pi \delta}{Jr}\Big).
\label{eq:RectSaw}
\end{align}

The saw-tooth model looks a little artificial, 
but its properties are quite similar to 
the Rice-Mele model~\cite{RiceMele1982PRL}
\begin{align}
 \hat{\mathcal{H}}_{\mathrm{RM}}=\sum_{j}
   &[(-t+\gamma)(\hat{a}_{2j}^{\dagger}\hat{a}_{2j+1}
     +\hat{a}_{2j+1}^{\dagger}\hat{a}_{2j})
\nonumber\\
   &+(-t-\gamma)(\hat{a}_{2j+1}^{\dagger}\hat{a}_{2j+2}
     +\hat{a}_{2j+2}^{\dagger}\hat{a}_{2j+1})
\nonumber\\
   &-\delta \hat{a}_{2j}^{\dagger}\hat{a}_{2j}
     +\delta \hat{a}_{2j+1}^{\dagger}\hat{a}_{2j+1}],
\end{align}
which is a well-known model for topological 
insulators~\cite{Nakajima2016NatPhys,WangTroyerDai2013PRL}. 
The distinction between the saw-tooth and Rice?Mele models is that, 
in the former, a pair of sites share the same position along the $x$ axis, 
whereas in the latter, each site occupies a different position. 
The Rice-Mele model was analyzed by a nonreciprocal LZ model 
in Ref.~\cite{Kitamura2020CommPhys}, and its shift current is studied 
by Floquet theory and Keldysh method~\cite{Morimoto2016SciAdv}. 

We also note that the 1D lattice electron model 
can be mapped to the corresponding $S=1/2$ spin chain model 
through the Jordan-Wigner transformation: 
\begin{align}
\begin{split}
 &\hat{S}_{l}^{z}=\hat{n}_{l}-\frac{1}{2},\\
 &\hat{S}_{l}^{+}=\hat{a}_{l}^{\dagger}\prod_{j=1}^{l-1}(1-2\hat{n}_{j}),\quad
 \hat{S}_{l}^{-}=\hat{a}_{l}\prod_{j=1}^{l-1}(1-2\hat{n}_{j}),
\end{split}
\end{align}
where $\hat{n}_{j}\equiv \hat{a}_{j}^{\dagger}\hat{a}_{j}$. 
Using this transformation, we cap map the saw-tooth model 
Eq.~\eqref{eq:HSaw} into a spin chain model 
with bond alternation and staggered field:
\begin{align}
 \hat{\mathcal{H}}
   =\sum_{j}
   [&-J(\hat{S}_{j,\mathrm{A}}^{x}\hat{S}_{j,\mathrm{B}}^{x}
       +\hat{S}_{j,\mathrm{A}}^{y}\hat{S}_{j,\mathrm{B}}^{y})
\nonumber\\
     &-Jr(\hat{S}_{j,\mathrm{B}}^{x}\hat{S}_{j+1,\mathrm{A}}^{x}
        +\hat{S}_{j,\mathrm{B}}^{y}\hat{S}_{j+1,\mathrm{A}}^{y})
\nonumber\\
    &-\delta \hat{S}_{j,\mathrm{A}}^{z}
     +\delta \hat{S}_{j,\mathrm{B}}^{z}].
\end{align}
In this mapping, the electric field corresponds 
to magnetic field gradient along the $S^{z}$ axis, 
and electric current corresponds to spin current. 
Multiferroic materials have been studied by 
the effective spin models, 
and the generation of electric shift current 
by electromagnons~\cite{Morimoto2019PRB} 
and spin shift current~\cite{Ishizuka2019PRL} 
have been proposed theoretically. 
The spin version of Rice-Mele model and spin pumping 
is discussed in Ref.~\cite{Shindou2005JPSJ}.

\subsection{Dirac systems}

The TLZ model can also be utilized for analyzing 
the Dirac systems with the application of circularly polarized laser field. 
The geometric tunneling in the Dirac systems is 
thoroughly studied in Ref.~\cite{Takayoshi2021SciPost}, 
and we briefly review the results here. 

First, let us consider the 2D Dirac fermions 
in the rotating electric field 
$\boldsymbol{E}=E(\cos(\Omega t),\sin(\Omega t))$ 
$(E>0)$. 
The vector potential is represented as 
$\boldsymbol{A}=A(-\sin(\Omega t),\cos(\Omega t))$, 
where $A=E/\Omega$ 
($\boldsymbol{E}=-\partial_{t}\boldsymbol{A}$). 
Then the Hamiltonian is written as 
\begin{align}
 \hat{\mathcal{H}}
   =v[\xi(k_{x}+e_{0}A_{x})\hat{\sigma}^{x}
     +(k_{y}+e_{0}A_{y})\hat{\sigma}^{y}]
     +m\hat{\sigma}^{z},
 \label{eq:laserHamiltonian}
\end{align}
where $v$ is the light velocity, which corresponds to 
the Fermi velocity in condensed matter, 
and $m>0$ is the mass. 
Note that there are two kinds of chirality $\xi=\pm$. 

%%%%%%%%%% Fig : Tunneling probability %%%%%%%%%%
\begin{figure}[t]
\centering
\includegraphics[width=0.48\textwidth]{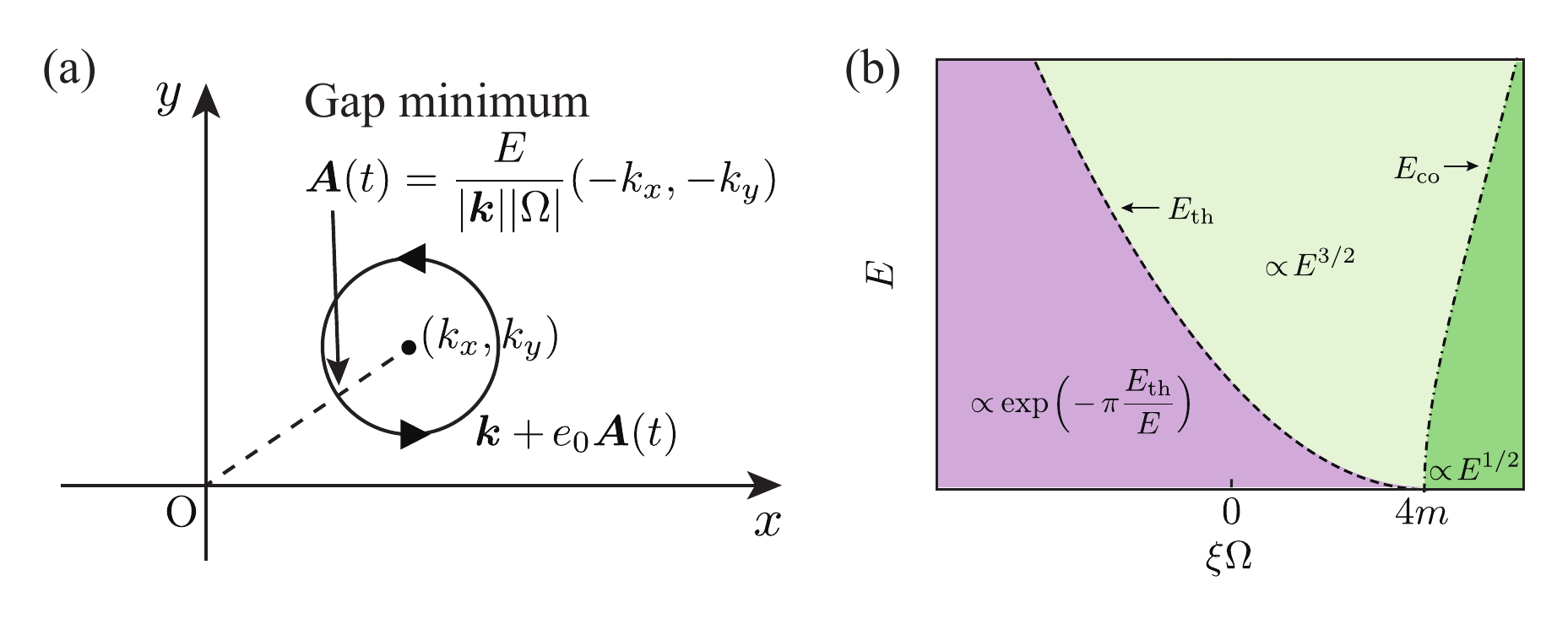}
\caption{(a) The gap minimum point in the wave number space. 
(b) The behavior of the total pair production rate $\Gamma_{\xi}$
in the parameter space of $\xi\Omega$ and $E$.}
\label{fig:Dirac2D}
\end{figure}
%%%%%%%%%%%%%%%%%%%%

For the fixed wave number $(k_{x},k_{y})$, 
we can regard Eq.~\eqref{eq:laserHamiltonian} as the TLZ model 
through the expansion up to the second order of $\Omega t$ 
around the gap minimum point 
$\big(\cos(\Omega t_{0}),\sin(\Omega t_{0})\big)
=\frac{\Omega}{|\boldsymbol{k}||\Omega|}(-k_{y},k_{x})$ 
[See Fig.~\ref{fig:Dirac2D}(a)], 
the Hamiltonian is approximated as 
\begin{align}
 \hat{\mathcal{H}}
   =&v\xi\Big(1-\frac{e_{0}E}{|\boldsymbol{k}||\Omega|}
     \Big)k_{x}\hat{\sigma}^{x}
   +v\Big(1-\frac{e_{0}E}{|\boldsymbol{k}||\Omega|}
     \Big)k_{y}\hat{\sigma}^{y}
   +m\hat{\sigma}^{z}
\nonumber\\
   &-v\frac{e_{0}E}{|\boldsymbol{k}||\Omega|}
     (\xi k_{y}\hat{\sigma}^{x}-k_{x}\hat{\sigma}^{y})
     (-\Omega t')
\nonumber\\
   &+v\frac{e_{0}E}{|\boldsymbol{k}||\Omega|}
     (\xi k_{x}\hat{\sigma}^{x}+k_{y}\hat{\sigma}^{y})
     \frac{(\Omega t')^{2}}{2},
\end{align}
where $t'=t-t_{0}$. 
Note that the angular frequency $\Omega$ works as 
a sweep velocity $F$ in the present case. 
The sign of $\Omega$ corresponds to the helicity of light. 
Utilizing Eq.~\eqref{eq:TLZgeneral}, 
we can calculate the tunneling probability as 
\begin{align}
\mathcal{P}_{\xi}(\Omega,\boldsymbol{k})
   =\exp\Bigg[-\pi
   \frac{\displaystyle\Big(M-\frac{\xi\Omega m}{4M}\Big)^{2}}
     {ve_{0}E}\Bigg],
\label{eq:P_MonoGra}
\end{align}
where we have defined 
$M=\sqrt{v^{2}(|\boldsymbol{k}|-e_{0}E/(|\Omega|))^{2}+m^{2}}$. 
Note that the tunneling probability Eq.~\eqref{eq:P_MonoGra} 
depends on both the chirality $\xi$ 
and the helicity of light $\mathrm{sgn}(\Omega)$. 

We can observe the characteristic phenomena 
caused by the geometric effects: 
the chiral imbalance and circular dichroism, 
which are represented as 
\begin{align}
 \frac{\mathcal{P}_{+}(\Omega,\boldsymbol{k})}
         {\mathcal{P}_{-}(\Omega,\boldsymbol{k})}
   =\exp\Big(\frac{\pi\Omega m}{ve_{0}E}\Big)
\label{eq:Imbalance}
\end{align}
and 
\begin{align}
 \frac{\mathcal{P}_{\xi}(|\Omega|,\boldsymbol{k})}
         {\mathcal{P}_{\xi}(-|\Omega|,\boldsymbol{k})}
   =\exp\Big(\frac{\pi\xi |\Omega| m}{ve_{0}E}\Big).
\label{eq:Dichro}
\end{align}
These results are analogous to the rectification, 
which has seen in the saw-tooth chain 
Eq.~\eqref{eq:RectSaw}. 
Since the ratio Eqs.~\eqref{eq:Imbalance} and \eqref{eq:Dichro}
does not depend on $\boldsymbol{k}$, 
the same relation holds 
for the fermion-antifermion pair production rate
\begin{align}
 \Gamma_{\xi}(\Omega) \equiv
   \frac{|\Omega|}{(2\pi)^{3}} \int d\boldsymbol{k}
     \mathcal{P}_{\xi}(\Omega,\boldsymbol{k}).
\end{align}

Another characteristic geometric effect, 
perfect tunneling, also plays an important role 
in fermion-antifermion pair production rate 
and demonstrates the twisted Schwinger effect 
From Eq.~\eqref{eq:P_MonoGra}, 
the condition for perfect tunneling 
$\mathcal{P}_{\xi}(\Omega,\boldsymbol{k})=1$ is 
$M=\frac{\xi\Omega m}{4M}$, 
and the solution of this equation is given as
\begin{align}
 |\boldsymbol{k}|=
\begin{cases}
\displaystyle
\mathrm{No \; solution}
&(\xi\Omega<4m)
\\
\displaystyle
\frac{e_{0}E}{|\Omega|}+
   \frac{1}{v}\sqrt{m\Big(\frac{\xi\Omega}{4}-m\Big)}
&(\xi\Omega\geq 4m, E<E_{\mathrm{co}})
\\
\displaystyle
\frac{e_{0}E}{|\Omega|}\pm
   \frac{1}{v}\sqrt{m\Big(\frac{\xi\Omega}{4}-m\Big)}
&(\xi\Omega\geq 4m, E\geq E_{\mathrm{co}}),
\end{cases}
   \label{eq:ptwn}
\end{align}
where 
$E_{\mathrm{co}}=\frac{|\Omega|}{e_{0}v}
\sqrt{m\big(\frac{\xi\Omega}{4}-m\big)}$. 
The contribution to pair production rate $\Gamma_{\xi}(\Omega)$ 
comes mainly from the wave numbers 
where perfect tunneling happens. 
In $\xi\Omega<4m$, the contribution is from 
the modified gap minimum $|\boldsymbol{k}|=e_{0}E/|\Omega|$. 

The pair production rate $\Gamma_{\xi}(\Omega)$ 
can be evaluated by the saddle point method~\cite{Takayoshi2021SciPost}. 
Here we only describe the results. 
For $\xi\Omega\geq 4m$, $\Gamma_{\xi}(\Omega)$ behaves 
in a power law of $E$. 
As $E$ is increased, the power shows a crossover 
from $\Gamma_{\xi}(\Omega)\propto E^{1/2}$ 
to $\Gamma_{\xi}(\Omega)\propto E^{3/2}$
at $E_{\mathrm{co}}$. 
For $\xi\Omega< 4m$, 
the pair production rate is written as 
\begin{align}
 \Gamma_{\xi}\propto E^{3/2}
   \exp\Big(-\pi\frac{E_{\mathrm{th}}(\Omega)}{E}\Big),
\label{eq:Rate_Schwin} 
\end{align}
where the threshold of electric field is provided as 
\begin{align}
 E_{\mathrm{th}}(\Omega)
   =\frac{(m-\xi \Omega /4)^{2}}{ve_{0}}.
\label{eq:TwistSchwingerlimit}
\end{align}
For small $E$, the pair production is exponentially suppressed. 
When $E$ is increased, the number of fermion-antifermion pairs 
grows rapidly above $E_{\mathrm{th}}(\Omega)$, 
and it shows a power law behavior 
$\Gamma_{\xi}\propto E^{3/2}$. 
In the limit of $\Omega \to 0$, 
the threshold becomes 
$E_{\mathrm{th}}(\Omega) \to m^{2}/(ve_{0})$ 
and the results by Schwinger 
Eqs.~\eqref{eq:Schwin} and \eqref{eq:Schwinlim}
are reproduced. 
Hence, Eqs.~\eqref{eq:Rate_Schwin} and 
\eqref{eq:TwistSchwingerlimit} are an extension of 
Schwinger's results 
to the case of rotating electric field. 
The behavior of the total pair production rate $\Gamma_{\xi}$
in the parameter space of $\xi\Omega$ and $E$ 
is summarized in Fig.~\ref{fig:Dirac2D}(b). 
A resemblance is pointed out between 
Fig.~\ref{fig:Dirac2D}(b) and the phase diagram of 
a holographic model~\cite{Hashimoto2017JHEP,Kinoshita2018JHEP}.

In several kinds of 2D materials such as 
monolayer transition metal dichalcogenides and 
graphene~\cite{Rycerz2007NatPhys,Schaibley2016NatRevMater,Liu2019Nano}, 
the 2D massive Dirac model works as an effective model 
for the dispersion around $K$ and $K'$ points, 
which are called valleys. 
These different valleys correspond to different chirality 
in the Dirac model. 
In the above discussion, 
it is assumed that the Fermi energy is located at zero.

The twisted Schwinger effect also appears 
in 3D massless Dirac fermions in circularly polarized laser, 
which is represented 
by the Hamiltonian 
\begin{align}
 \hat{\mathcal{H}}_{\mathrm{3D}}
   =v\sum_{j=x,y,z}
     \hat{\gamma}^{0}\hat{\gamma}^{j}(k_{j}+e_{0}A_{j}),
 \label{eq:laserHam3D}
\end{align}
$\boldsymbol{A}=A(-\sin(\Omega t),\cos(\Omega t),0)$. 
This Hamiltonian is the same as Eq.~\eqref{eq:Dirac3D}, 
and we set $m=0$ and substitute $k_{j}$ by $k_{j}+e_{0}A_{j}$. 
The Hilbert space of the 3D massless Dirac model Eq.~\eqref{eq:laserHam3D} 
is divided into two chiral sectors as 
\begin{align}
 \hat{\mathcal{H}}_{\mathrm{3D}}=
\begin{pmatrix}
 \hat{\mathcal{H}}_{-} & 0 \\
 0 & \hat{\mathcal{H}}_{+}
\end{pmatrix},
\end{align}
where 
$\hat{\mathcal{H}}_{\pm}=\pm v
\sum_{j=x,y,z}(k_{j}+e_{0}A_{j})\hat{\sigma}^{j}$.
Through the rotation of $\hat{\mathcal{H}}_{-}$ about 
the $\hat{\sigma}^{z}$ axis by the angle $\pi$, 
the Hamiltonian transforms into 
\begin{align}
 \hat{U}^{\dagger}\hat{\mathcal{H}}_{\mathrm{3D}}\hat{U}=
\begin{pmatrix}
 \hat{\mathcal{H}}_{-}' & 0 \\
 0 & \hat{\mathcal{H}}_{+}'
\end{pmatrix}
,\quad
 \hat{U}=
\begin{pmatrix}
 e^{i\frac{\pi}{2}\hat{\sigma}^{x}} & 0 \\
 0 & I
\end{pmatrix}.
\end{align}
The Hamiltonian in each sector 
$\hat{\mathcal{H}}_{+}'$ and $\hat{\mathcal{H}}_{-}'$ 
is the same as the massive 2D Dirac model 
Eq.~\eqref{eq:laserHamiltonian} 
with the replacement $m \to k_{z}$. 
Thus we can analyze the massless 3D Dirac model 
in the same way as the 2D case. 
We remark that the summation over $k_{z}$ and $\xi=\pm$ 
should be taken in order to evaluate the physical quantities. 
For example, the pair production rate 
and the current along the $z$ axis in each chiral sector 
are defined as
\begin{align}
 &\Gamma_{\xi}(\Omega)
   =\frac{|\Omega|^{4}}{(2\pi)^{4}v^{3}}
     \int d \boldsymbol{k}
     \tilde{\mathcal{P}}_{\xi}(\Omega,\boldsymbol{k})
\\
 &J_{\xi}^{z}(\Omega)
   =-\frac{2e\tau |\Omega|^{4}}{(2\pi)^{4}v^{2}}
     \int d \boldsymbol{k}
     \frac{k_{z}}{|\boldsymbol{k}|}
     \tilde{\mathcal{P}}_{\xi}(\Omega,\boldsymbol{k}),
\end{align}
where $\tau$ is the lifetime of the generated pairs 
and $\tilde{\mathcal{P}}_{\xi}(\Omega,\boldsymbol{k})$ is 
obtained by replacing $m$ in Eq.~\eqref{eq:P_MonoGra} by $k_{z}$. 
Since there is a symmetry 
$\tilde{\mathcal{P}}_{-\xi}(\Omega,k_{x},k_{y},k_{z})
=\tilde{\mathcal{P}}_{\xi}(\Omega,k_{x},k_{y},-k_{z})$ 
from Eq.~\eqref{eq:P_MonoGra},
the chiral pair production rate 
$\Gamma_{+}(\Omega)-\Gamma_{-}(\Omega)$ and 
the total current
$J_{+}^{z}(\Omega)+J_{-}^{z}(\Omega)$ 
becomes zero. 
Hence we focus on 
the total pair production rate 
$\Gamma_{+}(\Omega)+\Gamma_{-}(\Omega)$ and 
the chiral current
$J_{+}^{z}(\Omega)-J_{-}^{z}(\Omega)$. 
Similarly to the case of 2D Dirac models, 
the contributions from the perfect tunneling points are dominant. 
Since the integration over all the wave numbers 
$k_{x},k_{y},k_{z}$ is taken, 
the perfect tunneling points always exist 
in the integral region. 
Thus the total pair production rate and 
the chiral current behaves in the power law of 
the applied electric field $E$. 
In addition, the crossover of power is observed 
when $E$ is increased in the same way 
as the 2D case~\cite{Takayoshi2021SciPost}. 

The results explained above have an implication for the systems 
excited by a circularly polarized 
laser~\cite{Karch2010PRL,Wang2013Sci,Mciver2020NatPhys} and 
optical lattices with potential shaking~\cite{Jotzu2014Nat}. 
The chiral imbalance of pair production is related to 
the valley polarization~\cite{Yao2008PRB,Xiao2012PRL}. 
There are a lot of studies on other nonperturbative phenomena 
caused by circularly polarized laser 
fields~\cite{Oka2009PRB,Kitagawa2011PRB,Lindner2011NatPhys,
Wang2014EPL,Ebihara2016PRB,Chan2016PRL,Bucciantini2017PRB}.

\section{Keldysh crossover in many-body systems}
\label{sec:Crossover}

In the preceding discussion, 
we focused on quantum tunneling phenomena 
induced by the application of an external field. 
Another important approach for analyzing the dynamics 
is the Floquet theory, which is useful for 
describing multiphoton absorption processes 
in the high-frequency regime. 
A crossover occurs when the parameters cross the Keldysh line, 
which separates the tunneling regime 
from the Floquet regime (Fig.~\ref{fig:Keldysh}). 
This crossover, which is first studied by Keldysh 
in the ionization process of atoms~\cite{Keldysh1965JETP},
is therefore referred to as the Keldysh crossover. 
Following the seminal work by Keldysh, 
studies have been conducted on such crossover behavior 
in pair production under an AC electric 
field~\cite{Brezen1970PRD,Popov1974SovJNP,Popov2004PhysUsp}.
The Keldysh crossover is also studied in condensed matter systems 
such as Mott insulators under a laser field~\cite{Oka2012PRB}, 
and there are experimental observations 
in Mott insulators~\cite{Li2022PRL} 
and Weyl semimetals~\cite{Ikeda2024JPSJ} 
with an application of AC electric field. 

Here we consider the spin-1 Heisenberg antiferromagnetic chain, 
which has an excitation gap in contrast to the case of half-odd spin chains 
from Haldane's theory~\cite{Haldane1983PhysLett,Haldane1983PRL}. 
The ground state of the spin-1 chain has a topological order 
protected by the symmetry of the 
system~\cite{Gu2009PRB,Pollmann2010PRB,Pollmann2012PRB}. 
We apply a magnetic field gradient to 
the spin-1 Heisenberg antiferromagnetic chain, 
and then Hamiltonian is represented as 
\begin{align}
 \hat{\mathcal{H}}
   =&J\sum_{j}\Big[\frac{1}{2}(
     \hat{S}_{j}^{+}\hat{S}_{j+1}^{-}
     +\hat{S}_{j}^{-}\hat{S}_{j+1}^{+})
     +\hat{S}_{j}^{z}\hat{S}_{j+1}^{z}\Big]
\nonumber\\
   &-F\sum_{j}j\hat{S}_{j}^{z}.
\end{align}
Since $\sum_{j}\hat{S}_{j}^{z}$ is a conserved quantity 
for this Hamiltonian, 
$\hat{S}_{j}^{z}$ can be regarded as a charge at the site $j$. 
The $\hat{S}_{j}^{+}\hat{S}_{j+1}^{-}$ and 
$\hat{S}_{j}^{-}\hat{S}_{j+1}^{+}$ terms correspond to 
the hopping of the charge, 
and the $\hat{S}_{j}^{z}\hat{S}_{j+1}^{z}$ term 
corresponds to the nearest neighbor interaction. 
Since the $\sum_{j}j\hat{S}_{j}^{z}$ works as a polarization, 
we call $F$ the spin-electric field. 
In the velocity gauge, the Hamiltonian is represented as  
\begin{align}
 \hat{\mathcal{H}}(t)
   =&J\sum_{j}\Big[\frac{1}{2}(
     e^{iFt}\hat{S}_{j}^{+}\hat{S}_{j+1}^{-}
     +e^{-iFt}\hat{S}_{j}^{-}\hat{S}_{j+1}^{+})
     +\hat{S}_{j}^{z}\hat{S}_{j+1}^{z}\Big].
\end{align}
We consider the time evolution starting from 
the ground state of $\hat{\mathcal{H}}(0)$ 
in the time interval of $0\leq t\leq 2\pi/(LF)$, 
where $L$ is the system size. 
This process is the insertion of a single 
flux $\phi=2\pi/L$~\cite{Okazaki2025PRB}. 

Let us consider this situation 
from the perspective of a tunneling problem. 
The instantaneous energy level structure of 
$\hat{\mathcal{H}}\big(2\pi/(LF)\big)$ 
is the same as that of $\hat{\mathcal{H}}(0)$, 
and the excitation gap takes minimum at $t=\pi/(LF)$. 
Hence the time evolution in $0\leq t\leq 2\pi/(LF)$ 
can be regarded as tunneling phenomena, 
and the transition probability from the lower to upper band 
can be evaluated by the DDP method as 
\begin{align}
 P(F)=\exp(-\pi F_{\mathrm{th}}/F),
\label{eq:PTunnel}
\end{align}
where $F_{\mathrm{th}}$ is the threshold. 

This time evolution can also be analyzed 
in terms of the Floquet theory. 
Since the Hamiltonian has a periodicity 
$\hat{\mathcal{H}}(0)=\hat{\mathcal{H}}\big(2\pi/F\big)$, 
$F$ works as a angular frequency and we can define 
the Floquet effective Hamiltonian $\hat{H}_{\mathrm{F}}$ by 
\begin{align}
 \mathcal{T}
   \exp\bigg(-i\int_{t_{0}}^{t_{0}+2\pi/F}\hat{\mathcal{H}}(s)ds\bigg)
     =\exp(-i\hat{H}_{\mathrm{F}}t),
\end{align}
where $\mathcal{T}$ is the time-ordering. 
We can calculate $\hat{H}_{\mathrm{F}}$ using the inverse frequency 
expansion~\cite{Bukov2015AdvPhys,Mikami2016PRB,Oka2019AnnRev} 
and evaluate the transition probability 
from the lower to upper band as 
\begin{align}
 P(F)=c_{0}\exp(-\pi F_{q}/F),
\label{eq:Floquet}
\end{align}
where $c_{0}$ and $F_{q}$ are constant real numbers 
($0\leq c_{0}\leq 1$)~\cite{Okazaki2025PRB}. 

Comparing the numerically calculated $P(F)$ with the expressions 
Eqs.~\eqref{eq:PTunnel} and \eqref{eq:Floquet}, 
it is found that $P(F)$ is well described by 
Eq.~\eqref{eq:PTunnel} [Eq.~\eqref{eq:Floquet}]
in the region of $F<F_{*}$ [$F>F_{*}$], 
where $F_{*}$ is the crossing point of the two curves 
represented by Eqs.~\eqref{eq:PTunnel} 
and \eqref{eq:Floquet}~\cite{Okazaki2025PRB}. 
As $F$ is increased, the crossover from the tunneling regime 
to the Floquet regime happens at $F_{*}$. 
This crossover corresponds to the parameter change along 
the horizontal line in Fig.~\ref{fig:Keldysh} 
since $F$ serves as the angular frequency in the present case. 
Therefore this phenomenon is the Keldysh crossover 
in the quantum spin systems provoked by a DC spin electric field.

\section{Summary and discussion}
\label{sec:Summary}

We have discussed geometric effects of quantum tunneling, 
especially on the tunneling amplitude. 
After the review of the LZ model and Schwinger effect, 
the modern perspective on polarization, 
and the adiabatic perturbation theory, 
we have introduced the TLZ model and showed that 
the shift vector appears in the expression of tunneling probability. 
We have demonstrated several applications of the TLZ model, 
where nonreciprocity and perfect tunneling play an essential role. 
Lastly, we explain the Keldysh crossover seen in condensed matter systems. 

The tunneling phenomenon is closely related to the manipulation of 
quantum states. Hence developing the way to control quantum states 
rapidly and coherently by utilizing geometric effects in tunneling 
is an important challenge. 
In particular, geometric tunneling in strongly correlated systems 
still remains unexplored and should be further investigated 
in the future research. 
Furthermore, studying the interplay between geometric effects 
in quantum dynamics and the intrinsic topology of the system 
is also an intriguing future problem.

\begin{acknowledgment}
This work was supported by a Grant-in-Aid for 
Scientific Research from JSPS, KAKENHI Grant 
Nos. JP23H04865, JP23K22418, JP23K22487, JP24H00191, JP24K06891, 
and JST CREST Grant No. JPMJCR19T3, Japan.
\end{acknowledgment}

\end{document}